\documentclass[aps,prb,superscriptaddress,onecolumn,floatfix,notitlepage]{revtex4-2}
\usepackage[abs]{overpic}
\usepackage{graphicx,graphics}
\usepackage{dcolumn}
\usepackage{amsmath,amssymb,amsfonts}
\usepackage{wasysym}
\usepackage{latexsym,verbatim}
\usepackage{bm}
\usepackage{color}
\usepackage{ulem}
\usepackage[framemethod=default]{mdframed}
\usepackage[breaklinks=true,colorlinks,citecolor=blue,linkcolor=blue,urlcolor=blue]{hyperref}
\usepackage[makeroom]{cancel}
\usepackage{fancybox}
\usepackage{bbm}

\usepackage{soul}

\newcommand{\be}{\begin{eqnarray}}
\newcommand{\ee}{\end{eqnarray}}

\renewcommand{\Im}{\operatorname{\mathfrak{Im}}}

\begin{document}
\author{Alexander Kazantsev}
\email{alexander.kazantsev@manchester.ac.uk}
\affiliation{School of Physics and Astronomy, University of Manchester, M13 9PL, Manchester, UK}
\author{Alexey Berdyugin}
\affiliation{School of Physics and Astronomy, University of Manchester, M13 9PL, Manchester, UK}
\author{Andre Geim}
\affiliation{School of Physics and Astronomy, University of Manchester, M13 9PL, Manchester, UK}
\author{Alessandro Principi}
\email{alessandro.principi@manchester.ac.uk}
\affiliation{School of Physics and Astronomy, University of Manchester, M13 9PL, Manchester, UK}

%
\title{On the origin of Abrikosov’s quantum linear magnetoresistance}

\begin{abstract}

Compensated semimetals with Weyl spectra are predicted to exhibit unsaturated linear growth of their resistivity in quantizing magnetic fields. This so-called quantum linear magnetoresistance was introduced by Abrikosov, but approximations used in the theory remained poorly specified, often causing a confusion about experimental situations in which the analysis is applicable. Here we derive Abrikosov’s exact result using an alternative formalism based on diffusion of cyclotron orbits in a random potential. We show that both Weyl spectrum and a disorder smooth on the scale of the magnetic length are essential conditions for the validity of the theory, and the linear magnetoresistance appears in the extreme quantum limit where only the zeroth Landau level is half filled. It is the interplay between the relativistic-like nature of Weyl fermions and the classical dynamics of their cyclotron centers, which leads to the linear magnetoresistance. We also derive an analogous result in two dimensions, which has been missing in the literature and is relevant for numerous graphene-based systems.  
 
\end{abstract}

\maketitle

\section*{Introduction}

In Ref.~\cite{Abrikosov}, Abrikosov suggested the model of a compensated Weyl semimetal to explain the linear magnetoresistance in Ref. \cite{Xu1997} in the extreme quantum limit, such that $kT\ll \hbar \omega_c$ and $\omega_c\tau\gg 1$, where $k$ is the Boltzmann constant, $T$ is the temperature, $\omega_c$ is the cyclotron frequency and $\tau$ is the elastic scattering time. This result has often been quoted as an explanation for the linear magnetoresistance of three- and two-dimensional semimetals  (see, for example,  Refs.~\cite{Friedman, PhysRevB.85.041101, doi:10.1021/acs.nanolett.1c01647}) and the linear magnetoresistance itself was in turn interpreted as a sign of linear band dispersion even through in some cases the studied experimental systems exhibiting linear magnetoresistance were doped with several Landau levels occupied, clearly beyond the approximations used by Abrikosov.   In his original paper, Abrikosov used the formalism of Feynman diagrams in Matsubara imaginary time  to calculate the resistivity tensor, while also assuming that the main scatterers were screened ionized impurities. The technique that Abrikosov used somewhat obscured the physical picture behind the linear magnetoresistance. For instance, it makes it look as though the conductivity emerges due to transitions between Landau levels $0$ and $\pm 1$, which in the extreme quantum limit appears counter-intuitive because one would think that in this case all dynamics is restricted to the zeroth Landau level. There are also a number of questions the original derivation does not give the answer to. One of them is to what extent Abrikosov's result applies to two-dimensional semimetals. Its applicability is not entirely obvious because an electron scattering on a screened ionized impurity in three dimensions will have a scattering amplitude and a phase space to scatter into which are vastly different from those in two dimensions.  

At the same time, there is a simpler and more intuitive way of calculating magnetoresistivity in the extreme quantum limit, namely, Kubo's center migration theory \cite{Kubo}, which treats the current as a result of diffusive motion of cyclotron centers in an external electric field. The formalism is technically equivalent to Abrikosov's theory but has a clear advantage of allowing one to naturally restrict the electron dynamics to a single Landau level. When applied to the calculation of magnetoresistivity of a Weyl semimetal, this method makes it clear that for the linear magnetoresistance to emerge the scattering potential has to be smooth on the scale of $l=\sqrt{c\hbar/eH}$, the magnetic length  (where $H$ is the applied magnetic field). In this case one can treat the dynamics of electrons in a classical way and linear magnetoresistance then manifests itself as a purely semiclassical effect, resulting from the interplay between the classical dynamics  of orbit centers and the linear-in-$H$ quantum density of states.   

The outline of this paper is as follows. In Section I  we calculate the magnetoconductivity of three-dimensional Weyl fermions using the quantum-mechanical version of Kubo's center migration theory. In Section II, taking advantage of the smoothness of the scattering potential, we also provide the semiclassical treatment of the same problem. In Section III, we point out the two-dimensional analog of Abrikosov's linear magnetoresistance. In Section IV we demonstrate the equivalence between the center migration theory and the Feynman diagram calculations.    

\section{Quantum-mechanical treatment}

Throughout the paper we set $\hbar=1$ and restore it in the final equations. Following Abrikosov,  we assume that the energy spectrum can be described by a three-dimensional Weyl fermion with the one-body Hamiltonian in the magnetic field in the form

\be \label{eq:H_0}
{\cal H}_0=v\bm{\sigma}\cdot \left(\bm{p}-\frac{e}{c}\bm{A}\right),
\ee

\noindent where $v$ is the Fermi velocity, $\bm{\sigma}=(\sigma_x, \sigma_y, \sigma_z)$ is  a vector of Pauli matrices, $\bm{A}$ is the vector potential, $e$ is the  electron charge and $c$ is the speed of light. Assume that a magnetic field of magnitude $H$ is applied along the $z$ axis. Then in Landau gauge $\bm{A}=(0, H x, 0)$ and the eigenstates of the Hamiltonian can also be chosen as eigenstates of momenta $p_y$ and $p_z$.  Below, instead of $p_y$, we will use $X=p_yl^2$ to label the stationary states. The stationary states are also labelled by a third quantum number $N$ taking arbitrary (positive and negative) integer values such that the energy levels have the form  $E_N(p_z)=\mbox{sign}(N)v\sqrt{p_z^2+2|N|/l^2}$ for $N\neq 0$ and $E_0(p_z)=vp_z$. The stationary state wavefunctions have the form

\be
|NXp_z\rangle=\begin{pmatrix}
\frac{1}{\sqrt{2}}\left(1+\frac{vp_z}{E_N}\right)^{1/2}h_{|N|}(x-X)\\
-\frac{i}{\sqrt{2}}\mbox{sign}(N)\left(1-\frac{vp_z}{E_N}\right)^{1/2}h_{|N|-1}(x-X)
\end{pmatrix}\frac{e^{iyX/l^2+ip_z z}}{\sqrt{L_yL_z}}
\ee

\noindent for $N\neq 0$ and 

\be 
|0Xp_z\rangle =\begin{pmatrix}
h_{0}(x-X)\\ 0
\end{pmatrix}\frac{e^{iyX/l^2+ip_z z}}{\sqrt{L_yL_z}}
\ee

\noindent for $N=0$, where $h_{|N|}(x)=(2^{|N|}|N|!\sqrt{\pi}l)^{-1/2}H_{|N|}(x/l)\exp[-x^2/(2l^2)]$ is the level $|N|$ quantum oscillator stationary state, with $H_{|N|}(x/l)$ the Hermite polynomial.  In Landau gauge, the stationary states in each Landau level are localized in the planes $x=X$, where $X$ is the cyclotron center position along $x$. For a finite sample with dimensions $L_x\times L_y\times L_z$, the quantum number $X$ takes values between $0$ and $L_x$  in steps of $2\pi l^2/L_y$. Since the energy levels do not depend on $X$, the degeneracy of each Landau level equals $L_x/(2\pi l^2/L_y)=L_xL_y/(2\pi l^2)$. One can also define the cyclotron center position along $y$ as $Y=y-l^2p_x$, but $[X,Y]=-il^2$ so $X$ and $Y$ do not commute and stationary states cannot be eigenstates of both.  

When the semimetal is fully compensated, {\it i.e.}, the densities of conduction electrons and holes are equal to each other and the chemical potential $\mu$ vanishes, the Hall conductivity $\sigma_{xy}$ vanishes too. Below we will assume full compensation and will be interested in calculating $\sigma_{xx}$. In the limit of high enough magnetic field, such that $kT\ll \omega_c$, where $\omega_c=\sqrt{2}v/l$, the carriers are mostly confined to the zeroth Landau level and, being localized around their cyclotron centers, have zero kinetic energy of motion in the $xy$ plane. When an electric field is applied along $x$, the carriers will start moving along $y$, which will result in a Hall current, vanishing when the densities of electrons and holes are equal to each other. To make electrons move along the electric field, one has to introduce scatterers that will make electrons hop between cyclotron centers performing a random walk. Then the longitudinal conductivity $\sigma_{xx}$ is related to the diffusion coefficient $D_{xx}$ associated with this random walk through Einstein relation

\begin{equation}\label{eq:einstein}
\sigma_{xx}=e^2\left(\frac{\partial n}{\partial \mu}\right)D_{xx}, 
\end{equation}  

\noindent where 

\begin{equation}\label{eq:comp}
\frac{\partial n}{\partial \mu}=\int d\varepsilon \nu(\varepsilon) \left(-\frac{\partial f}{\partial \varepsilon}\right)
\end{equation}

\noindent with $f(\varepsilon)=\big\{\!\exp\big[(\varepsilon-\mu)/(kT)\big]+1\big\}^{-1}$ the Fermi--Dirac distribution  and $\nu(\varepsilon)$ the density of states per unit volume defined as

\be\label{eq:dos}
\nu(\varepsilon)=\frac{2}{V}\sum_{Xp_z}\delta\big[\varepsilon-E_0(p_z)\big],
\ee

\noindent where $V=L_zL_yL_z$ is the volume of the sample,  a factor of $2$ accounts for spin and we also took into account that the contribution of the other Landau levels to $\partial n/\partial \mu$ apart form $N=0$ is exponentially small.  The diffusion coefficient equals

\be \label{eq:diffusion}
D_{xx}=\frac{1}{2}\sum_{p_z'X'}\frac{(X'-X)^2}{\tau_{Xp_z\to X'p_z'}},
\ee

\noindent where $\tau_{Xp_z\to X'p_z'}$ is the inverse transition rate between states $|0Xp_z\rangle$  and $|0X'p_z'\rangle$ (here we also neglected scattering between different Landau levels). When the scattering events are rare, {\it i.e.}, $\omega_c\tau\gg 1$ one can use the Fermi golden rule to evaluate the transition rate

\be \label{eq:rate}
\frac{1}{\tau_{Xp_z\to X'p_z'}}=2\pi\left\langle|\langle 0X'p_z'|U|0Xp_z\rangle|^2\right\rangle_s\delta\big[E_0(p_z)-E_0(p_z')\big],
\ee

\noindent where $U$ is the scattering potential and $\langle\dots\rangle_s$ stands for averaging over the positions of scatterers. If we substitute Eqs. \eqref{eq:comp}--\eqref{eq:rate} into Eq. \eqref{eq:einstein} we will obtain

\begin{equation}\label{eq:master}
\sigma_{xx}=\frac{2 e^2}{V}\sum_{Xp_z}\sum_{X'p_z'}\left(-\frac{\partial f}{\partial \varepsilon}\right)\bigg|_{\varepsilon=E_{0}(p_z)} \frac{(X-X')^2}{2}2\pi\Big\langle\big|\langle 0Xp_z|U|0X'p_z'\rangle\big|^2\Big\rangle_s\delta\big[E_0(p_z)-E_{0}(p_z')\big],
\end{equation}

\noindent which is one of the central equations of the center migration theory, see \cite{Kubo}. Let us now evaluate the diffusion coefficient given by Eqs. \eqref{eq:diffusion}--\eqref{eq:rate}. The scattering potential $U$ is a sum of potentials created by different impurities

\begin{equation}
U(\bm{r})=\sum_{\bm{R}_i}u(\bm{r}-\bm{R}_i), 
\end{equation} 

\noindent where $u(\bm{r})=(1/V)\sum_{\bm{q}}u_{\bm{q}}\exp(i\bm{q}\bm{r})$ is the potential created by an impurity sitting at the origin and $\bm{R}_i$ is the position of each impurity. Assuming uniform impurity distribution and keeping only the term linear in the impurity concentration we obtain

\begin{equation}
\big\langle U(\bm{r})U(\bm{r}') \big\rangle_s=\frac{N_s}{V}\sum_{\bm{q}}|u_{\bm{q}}|^2e^{i\bm{q}(\bm{r}-\bm{r}')},
\end{equation}

\noindent where $N_s$ is the impurity concentration.  Let us evaluate the square of the transition matrix (recall that $X=p_yl^2$)

\begin{eqnarray}
\Big\langle\big|\langle 0Xp_z|U|0X'p_z'\rangle\big|^2\Big\rangle_s=&&\frac{N_s}{V}\sum_{\bm{q}}|u_{\bm{q}}|^2|\langle 0Xp_z|e^{i\bm{q}{r}}|0X'p_z'\rangle|^2\nonumber\\&&=\frac{N_s}{V}\sum_{\bm{q}}|u_{\bm{q}}|^2\delta_{X'-X+q_yl^2,0}\delta_{p_z'-p_z+q_z,0}\left|\frac{1}{\sqrt{\pi}l}\int dx e^{-(x-X)^2/(2l^2)}e^{iq_xx}e^{-(x-X')^2/(2l^2)}\right|^2\nonumber\\
&&=\frac{N_s}{V}\sum_{\bm{q}}|u_{\bm{q}}|^2\delta_{X'-X+q_yl^2,0}\delta_{p_z'-p_z+q_z,0}e^{-q_{\perp}^2l^2/2},
\end{eqnarray}

\noindent where $q_\perp^2=q_x^2+q_y^2$. Substituting this back into Eqs. \eqref{eq:diffusion}--\eqref{eq:rate} and performing summation with respect to $X'$ and $p_z'$ we obtain

\begin{equation}\label{eq:intermediate}
D_{xx}=\frac{1}{V}\sum_{\bm{q}}\frac{(q_yl^2)^2}{2}2\pi N_s |u_{\bm{q}}|^2e^{-q_\perp^2l^2/2}\delta\big[E_0(p_z)-E_0(p_z-q_z)\big]. 
\end{equation}

\noindent Let us change in this equation summation with respect to $\bm{q}$ to integration according to the rule 

\be
\sum_{\bm{q}}\dots= V\int \frac{d^3q}{(2\pi)^3}\dots
\ee
 
\noindent and also substitute $E_0(p_z)=vp_z$. This results in

\be\label{eq:diff_final}
D_{xx}=\int\frac{d^3q}{(2\pi)^3}\frac{(q_yl^2)^2}{2}2\pi N_s |u_{\bm{q}}|^2e^{-q_\perp^2l^2/2}\delta(vq_z)=\frac{1}{v}\int\frac{d^2q_\perp}{(2\pi)^2}\frac{(q_yl^2)^2}{2}N_se^{-q_\perp^2l^2/2}|u_{\bm{q}}|^2\big|_{q_z=0}.
\ee

\noindent The compressibility $\partial n/\partial\mu$, given by Eqs. \eqref{eq:comp}--\eqref{eq:dos}, equals at $\mu=0$

\be
\frac{\partial n}{\partial\mu}=\frac{2}{V}\sum_{Xp_z} \left(-\frac{\partial f}{\partial \varepsilon}\right)\bigg|_{\varepsilon=E_{0}(p_z)}=\frac{2}{(2\pi l^2)L_x}\int_0^{L_x} dX\int \frac{dp_z}{(2\pi)}\left(-\frac{\partial f}{\partial \varepsilon}\right)\bigg|_{\varepsilon=E_{0}(p_z)},
\ee

\noindent where we changed summation with respect to $X$ and $p_z$ to integration according to the rules

\begin{eqnarray}
\sum_{X}\dots=\frac{L_y}{2\pi l^2}\int_0^{L_x} dX\dots;\qquad \sum_{p_z}\dots=L_z\int\frac{dp_z}{2\pi}\dots
\end{eqnarray}

\noindent Performing the trivial integral with respect to $X$, we obtain 

\be\label{eq:comp_final}
\frac{\partial n}{\partial \mu}=\frac{1}{2\pi l^2} 2\int\frac{dp_z}{2\pi}\left(-\frac{\partial f}{\partial \varepsilon}\right)\bigg|_{\varepsilon=E_0(p_z)}=\frac{2}{2\pi l^2}\int\frac{dp_z}{2\pi kT}\frac{1}{(e^{vp_z/(2kT)}+e^{-vp_z/(2kT)})^2}=\frac{2}{(2\pi)^2l^2 v}.
\ee

\noindent Plugging Eqs. \eqref{eq:comp_final} and \eqref{eq:diff_final} into Eq. \eqref{eq:einstein}, we obtain

\begin{equation}\label{eq:semifinal}
\sigma_{xx}=\frac{2 e^2}{(2\pi)^2l^2v^2}\int\frac{d^2q_\perp}{(2\pi)^2}\frac{(q_yl^2)^2}{2}N_se^{-q^2_\perp l^2/2}|u_{\bm{q}}|^2\big|_{q_z=0}.
\end{equation}

\noindent Note that, if the potential is sufficiently smooth, that is to say if $|u_{\bm{q}}|^2\big|_{q_z=0}$ goes to zero faster than $\exp(-q_\perp^2l^2/2)$ as $q_\perp\to\infty$, then the exponential can be neglected in the integrand of Eq.  \eqref{eq:semifinal}. Then, the magnetic field dependence of $\sigma_{xx}$ derives from the factor $l^4$  due to the diffusion constant and from the factor $l^{-2}$ that came from the density of states so that in the end $\sigma_{xx}\propto l^2\propto 1/H$ and $\rho_{xx}=1/\sigma_{xx}\propto H$.  To see this in more detail let us, following  Abrikosov,  take charged partially screened impurities as primary scatterers. Then

\begin{equation}
u_{\bm{q}}=\frac{4\pi e^2}{\epsilon_\infty(q^2+\kappa^2)},
\end{equation}

\noindent where $q^2=q_\perp^2+q_z^2$ is transmitted momentum squared, $\epsilon_\infty$ is the effective dielectric constant and $\kappa^2$ is the Thomas-Fermi inverse screening length given by the equation

\begin{equation}\label{eq:scr_length}
\kappa^2=\frac{4\pi e^2}{\epsilon_\infty} \frac{\partial n}{\partial \mu}=\frac{2e^2}{\pi\epsilon_\infty v l^2}.
\end{equation}

\noindent If $e^2/(\epsilon_\infty v) \ll 1$ then $\kappa^2\ll 1/l^2$. Let us evaluate Eq. \eqref{eq:semifinal}. Note that $q_y^2$ there can be changed to $q_\perp^2/2$ due to cylindrical symmetry of the potential while $q_z=0$. We then obtain

\begin{equation}\label{eq:final}
\sigma_{xx}=2e^2l^2\left(\frac{e^2}{v\epsilon_\infty}\right)^2N_s\int\frac{d^2 q_\perp}{(2\pi)^2}\frac{q_\perp^2}{(q_\perp^2+\kappa^2)^2}e^{-q_\perp^2l^2/2}.
\end{equation}

\noindent Without the exponential, the integral in Eq. \eqref{eq:final} diverges logarithmically. The exponential then imposes a high-momentum cut-off on the order of $1/l$ on which the integral will depend logarithmically. On the other hand, the integral also diverges logarithmically at low momenta without $\kappa^2$ in the denominator. So $\kappa$  imposes a low-momentum cut-off, on which the integral will depend logarithmically too. Ultimately the value of the integral will be on the order of $\ln(\kappa l)$ and the dependence on $l$ will cancel between the low and the high momentum cut-off. Thus, to logarithmic accuracy, the integral evaluates to 

\begin{equation}
\sigma_{xx}=\frac{e^2}{2\pi\hbar}l^2\left(\frac{e^2}{\hbar v\epsilon_\infty}\right)^2N_s\ln\left(\frac{1}{\kappa^2 l^2}\right)=\frac{e^2}{2\pi\hbar}l^2\left(\frac{e^2}{\hbar v\epsilon_\infty}\right)^2N_s\ln\left(\frac{\epsilon_\infty v\hbar}{e^2}\right)\propto 1/H,
\end{equation}

\noindent where we also restored $\hbar$.  This is exactly Abrikosov's result~\cite{Abrikosov}.

\section{Semiclassical evaluation}

The same result could be obtained from classical dynamics.  Basically, what happens is this: when an electron passes an impurity with velocity $v$ along the magnetic field the center of its orbit is deflected in the plane perpendicular to the field. Because the motion along the magnetic field is unidirectional (see below) the electron never returns to the same impurity and localization effects can be neglected.  Let us consider the motion of an electron in the force field of an impurity more precisely.  Including the impurity potential in Eq.~(\ref{eq:H_0}), the Hamiltonian has the form 

\be
{\cal H}=v\bm{\sigma}\cdot \left(\bm{p}-\frac{e}{c}\bm{A}\right)+u(\bm{r}),
\ee

\noindent where $u(\bm{r})$ is the potential of a single impurity sitting at the origin. The coordinates $\xi_x$ and $\xi_y$ that describe the relative motion of the electron with respect to the cyclotron center are defined as

\be 
\xi_x=-l^2\left(p_y-\frac{e}{c}A_y\right);\qquad \xi_y=l^2\left(p_x-\frac{e}{c}A_x\right),
\ee

\noindent while the coordinates $X$ and $Y$ of the cyclotron center are defined as

\be 
X=x-\xi_x;\qquad Y=y-\xi_y.
\ee

\noindent The operators thus introduced have the following commutation relations among each other

\be
&&[\xi_x, \xi_y]=il^2;\quad[X, Y]=-il^2;\nonumber\\
&&[X,\xi_x]=[X,\xi_y]=[Y, \xi_x]=[Y,\xi_y]=0, 
\ee

\noindent and they all obviously commute with $z$ and $p_z$ and also pseudospin.   The equations of motion for $X$ and $Y$ read

\begin{equation}\label{eq:eom}
\dot{X}=i[{\cal{H}}, X]=l^2\partial_y u (\bm{r});\qquad \dot{Y}=i[{\cal{H}}, Y]=-l^2\partial_x u(\bm{r}), 
\end{equation}

\noindent where we took into account that $[X, y]=[X, Y+\xi_y]=-il^2$ and $[Y, x]=[Y, X+\xi_x]=il^2$. Let us restrict the dynamics to the zeroth Landau level. In this case it is not difficult to show that the uncertainty in the values of $\xi_x$ and $\xi_y$ reaches the value of $l/\sqrt{2}$. Therefore if the potential is smooth on the scale of $l$, one can neglect the difference between $x$ and $X$ and between $y$ and $Y$ on the right hand side of Eq. \eqref{eq:eom} and close the equations of motion with respect to $X$ and $Y$. On the other hand, as follows from the commutation relation between $X$ and $Y$, the uncertainty in the values of $X$ and $Y$ can be made as small as $l$. Therefore, for a smooth potential, instead of the Heisenberg equations of motion one can consider classical dynamics of point particles located at $\bm{R}=(X, Y)$. The equations of motion for them will have the form

\begin{eqnarray}\label{eq:class_eom}
\dot{X}=l^2\partial_Y u(\bm{R});\qquad
\dot {Y}=-l^2\partial_X u(\bm{R});\qquad
\dot{z}=\frac{\partial E_{0}(p_z)}{\partial p_z}=v,
\end{eqnarray} 

\noindent where $v$ is the Fermi velocity. For simplicity, let us assume the potential to be spherically symmetric. Then 

\begin{eqnarray}
&&\dot{X}=l^2(Y/\rho)\partial_\rho u;\\
&&\dot{Y}=-l^2(X/\rho)\partial_\rho u,
\end{eqnarray} 

\noindent where $\rho=\sqrt{X^2+Y^2}$ is the radial coordinate. If we multiply the first equation by $X$ and the second by $Y$ and take their sum we will get 

\begin{equation}
X\dot{X}+Y\dot{Y}=\frac{1}{2}\frac{d(\rho^2)}{dt}=0,
\end{equation}

\noindent which means that the radial coordinate remains the same. This is not surprising since we know that the drift of the cyclotron center must happen in the direction perpendicular to the force exerted by the impurity. So what happens is that the polar angle changes. Introducing the polar angle in the form

\begin{eqnarray}
&&X=\rho \cos\phi;\\
&&Y=\rho \sin \phi 
\end{eqnarray}

\noindent and substituting this into the equations of motion we obtain

\begin{eqnarray}
\dot{\phi}=-(l^2/\rho)\partial_\rho u.
\end{eqnarray}

\noindent Therefore the total change in the angle as the particle passes the impurity is equal to

\begin{equation}
\Delta\phi=-\int\limits_{-\infty}^{\infty}dt (l^2/\rho)\partial_{\rho} u.
\end{equation} 

\noindent Recall that $\dot{z}=v$, so that the integral with respect to $t$ can be changed to the integral with respect to $z$ in the following manner

\begin{equation}
\Delta \phi =-\frac{1}{v}\int\limits_{-\infty}^{\infty}dz (l^2/\rho)\partial_\rho u.
\end{equation}

Now imagine an electron incident on an impurity with an initial polar angle $\phi_1$. After it scatters off the impurity it will have a different polar angle $\phi_2=\phi_1+\Delta \phi$. The impact parameter $\rho$  ({\it i.e.} the distance from the impurity) will not change. The shift of the position of the center of the orbit is expressed then by the equations

\begin{eqnarray}
&&\Delta X=\rho(\cos \phi_2-\cos \phi_1);\\
&&\Delta Y=\rho(\sin \phi_2-\sin \phi_1).
\end{eqnarray}

\noindent On average, both shifts are equal to zero. But what will not be averaged to zero is the mean squared shift

\begin{equation}
(\Delta X)^2+(\Delta Y)^2=2\rho^2(1-(\cos \phi_2\cos\phi_1+\sin\phi_2\sin\phi_1))=2\rho^2(1-\cos(\phi_2-\phi_1))=2\rho^2(1-\cos\Delta \phi).
\end{equation}

\noindent Assuming $\Delta\phi\ll 1$ we obtain

\begin{equation}
(\Delta X)^2+(\Delta Y)^2=\rho^2(\Delta\phi)^2.
\end{equation}

At small impurity density, the scattering events on different impurities are uncorrelated and the total scattering rate equals $N_s v (d\sigma/d\rho)$ per unit impact parameter, where $d\sigma$ is the differential cross-section and $N_s$ is the impurity density, so the diffusion coefficient is then equal to 

\be\label{eq:class_diff}
D_{xx}=\frac{1}{2}(D_{xx}+D_{yy})&&{}=\frac{1}{2}\int d\rho N_s v \frac{d\sigma}{d\rho}\frac{(\Delta X)^2+(\Delta Y)^2}{2}\nonumber\\&&{}=\frac{1}{4}\int N_s v \;\frac{d \sigma}{d\rho} \rho^2(\Delta\phi)^2d\rho=\frac{1}{2}\pi N_s\int \rho\, d\rho\, v^{-1}l^4\left(\int\limits_{-\infty}^{+\infty}dz\partial_\rho u\right)^2,
\ee

\noindent where we just substituted $(d\sigma/d\rho)=2\pi\rho$.  This integral is the same as the one on the right hand side of Eq. \eqref{eq:semifinal}, only without the exponential $\exp[-q_\perp^2l^2/2]$ and written in the coordinate representation. The exponential is a sign of finite spread of the electron's wavefunction so it is not surprising that it does not appear in the classical treatment.  For completeness let us calculate the integral in Eq. \eqref{eq:class_diff} in the coordinate representation too. Take the screened Coulomb potential

\begin{equation}
u(\rho, z)=\frac{e^2 e^{-\kappa\sqrt{\rho^2+z^2}}}{\epsilon_\infty\sqrt{\rho^2+z^2}},
\end{equation}

\noindent where $\kappa$ is the inverse screening length calculated in Eq. \eqref{eq:scr_length} and $\epsilon_\infty$ is the effective dielectric constant. For $\rho$ greater than the screening length $1/\kappa$ the electron does not feel the scattering potential so that we can impose the upper limit on $\rho$ at $1/\kappa$ in the integral above and neglect the screening for $\rho<1/\kappa$. On the other hand, the characteristic length scale on which the unscreened Coulomb potential changes significantly is just the distance to the origin and in close proximity to the origin it changes very fast. Therefore, at distances to the origin smaller than $l$ the semiclassical approximation breaks down and therefore we have to impose a lower cut-off on $\rho$ at $l$ in Eq. \eqref{eq:class_diff}. We might of course have got away without it, if the integral were convergent at small values of $\rho$, but it is not, it is in fact logarithmically divergent. Thus the diffusion constant is equal to

\begin{eqnarray}
D_{xx}=&&\frac{1}{2}\pi N_s\int_{l}^{1/\kappa} \rho\, d\rho\, v^{-1}l^4\left(\int\limits_{-\infty}^{+\infty}dz \frac{e^2\rho}{\epsilon_\infty(\rho^2+z^2)^{3/2}}\right)^2\nonumber\\&&=2\pi N_s \frac{e^4 l^4}{\epsilon_\infty^2 v}\int\limits_{l}^{1/\kappa}\frac{d\rho}{\rho}=2\pi N_s \frac{e^4l^4}{\epsilon_\infty^2 v}\ln\left(\frac{1}{\kappa l}\right)=\pi N_s \frac{e^4l^4}{\epsilon_\infty^2 v}\ln\left(\frac{\epsilon_\infty v}{e^2}\right).
\end{eqnarray} 

\noindent Now, by Einstein relation

\begin{equation}
\sigma_{xx}=e^2(\partial n/\partial\mu)D_{xx}=e^2\frac{2}{(2\pi)^2l^2 v}\pi N_s \frac{e^4l^4}{\epsilon_\infty^2 v}\ln\left(\frac{\epsilon_\infty v}{e^2}\right)=\frac{e^2}{2\pi}l^2\left(\frac{e^2}{\epsilon_\infty v}\right)^2N_s\ln\left(\frac{\epsilon_\infty v}{e^2}\right),
\end{equation}

\noindent where $\partial n/\partial \mu$ was calculated in Eq. \eqref{eq:comp_final}.  Note again that the resulting dependence on $l$ comes from the interplay between the classical dynamics of Eq. \eqref{eq:class_eom} and the quantum-mechanical density of states in Eq. \eqref{eq:comp_final}. This result is exactly the same as the one given in Eq. \eqref{eq:final}.

\section{center migration in two dimensions}

In this section we will apply center migration theory to a two-dimensional semimetal in a smooth static disorder potential. Here we will be following the treatment  for  a non-relativistic electron gas presented in Refs. \cite{Kubo, Ando}. Consider a fully compensated graphene sheet ($\mu=0$) in a magnetic field $H$ applied perpendicularly to the plane of motion. The low energy excitations can be effectively described by four independent Dirac Hamiltonians, corresponding to two inequivalent valleys each with two opposite spin orientations. We will only consider a single Dirac fermion and multiply its contribution to the conductivity by four. The Hamiltonian for a two-dimensional Dirac fermion in magnetic field has the form

\be
{\cal H}_0=v\bm{\sigma}\cdot\left(\bm{p}-\frac{e}{c}\bm{A}\right),
\ee

\noindent where  now $\bm{\sigma}=(\sigma_x, \sigma_y)$. In Landau gauge $\bm{A}=(0, H x, 0)$ corresponding to  a magnetic field of magnitude $H$ applied along $z$. In Landau gauge the stationary states can be chosen as eigenstates of the operator $X$, and therefore they are labelled by values of their cyclotron center position along $x$ as well as by a Landau level number $N$ taking  (positive and negative) integer values. The energy levels are equal to $E_{N}=\mbox{sign}(N)\omega_c\sqrt{|N|}$, where $\omega_c=\sqrt{2}v/l$ and the stationary states are, for $N\neq 0$,

\begin{equation}
|NX\rangle=\begin{pmatrix}
\frac{1}{\sqrt{2}}h_{|N|}(x-X)\\
-\frac{i}{\sqrt{2}}\mbox{sign}(N)h_{|N|-1}(x-X)
\end{pmatrix}\frac{e^{iXy/l^2}}{\sqrt{L_y}}
\end{equation}

\noindent and, for $N=0$, 

\be
|0X\rangle=\begin{pmatrix}
h_{0}(x-X)\\0
\end{pmatrix}\frac{e^{iXy/l^2}}{\sqrt{L_y}}.
\ee
   
\noindent Here again $h_{|N|}(x-X)=(2^{|N|}|N|!\sqrt{\pi}l)^{-1/2}H_{|N|}\left[(x-X)/l\right]\exp[-(x-X)^2/(2l^2)]$ is the $|N|$-th energy level stationary state of a harmonic oscillator shifted by $X$ away from the origin, $L_y$ is the dimension of the sample along $y$, the dimension along $x$ being $L_x$.  The cyclotron center position $X$ takes values between $0$ and $L_x$ in steps of $2\pi l^2/L_y$.
We will again assume a smooth static disorder. This can be created by charged impurities residing outside the plane of motion, e.g. in the substrate. At temperatures satisfying $kT\ll \omega_c$ all the charge carriers (electrons and holes) reside in the zeroth Landau level so we can restrict all the dynamics to it.

We will again apply the central equation of the center migration theory, Eq. \eqref{eq:master}, only this time there is no quantum number $p_z$. This creates a problem because $\delta(E_0-E_0)$ is infinity. The root of the problem is that the unperturbed Landau levels are infinitely thin. To cure that we just have to take into account the broadening. This is achieved by the substitution

\be \label{eq:broadening}
\left(-\frac{\partial f}{\partial \varepsilon}\right)\bigg|_{\varepsilon=E_0}\delta(E_{0}-E_0)\to\int d\varepsilon \left(-\frac{\partial f}{\partial \varepsilon}\right)\langle S_{0X}(\varepsilon)\rangle_s\langle S_{0X'}(\varepsilon)\rangle_s,
\ee

\noindent where $S_{NX}$ is the spectral function of the $N$th Landau level defined as

\be
S_{NX}(\varepsilon)=\langle NX|\delta(\varepsilon -{\cal H})|NX\rangle=-\frac{1}{\pi}\Im G_{NX}(\varepsilon+i0),
\ee

\noindent where ${\cal H}={\cal H}_0+U(\bm{r})$ is the Dirac Hamiltonian perturbed by the disorder potential $U(\bm{r})$. In the equation above, $G_{NX}$ is the Green function of the electron in the $N$th Landau level in presence of disorder. 

Therefore, taking the broadening  in Eq.~(\ref{eq:broadening}) into account we obtain

\begin{equation}\label{eq:semi_master}
\sigma_{xx}=\frac{4 e^2}{V}\sum_{X}\sum_{X'}\int d\varepsilon\left(-\frac{\partial f}{\partial \varepsilon}\right)\frac{(X-X')^2}{2}2\pi\big\langle\big|\langle 0X|U|0X'\rangle\big|^2\big\rangle_s  \langle S_{0X}(\varepsilon)\rangle_s \langle S_{0X'}(\varepsilon)\rangle_s,
\end{equation}

\noindent where a factor of $4$ takes account of the spin and valley  degeneracy, and $V=L_xL_y$ is the area of the sample.

We will assume Gaussian disorder with correlation function

\begin{equation}
\langle U(\bm{r})U(\bm{r}'))\rangle_{s}=\mathcal{F}(\bm{r}-\bm{r'})=\frac{1}{V}\sum_{\bm{q}}\mathcal{F}_{\bm{q}}e^{i\bm{q}(\bm{r}-\bm{r}')},
\end{equation}

\noindent where

\be\label{eq:disorder_corr}
{\cal F}(\bm{r}-\bm{r}')=U_0^2\exp\left(-\frac{|\bm{r}-\bm{r}'|^2}{2\xi^2}\right);\qquad {\cal F}_{\bm{q}}=2\pi U_0^2\xi^2\exp\left(-\frac{\bm{q}^2\xi^2}{2}\right),
\ee

\noindent where $U_0$ is the characteristic strength and $\xi$ is the correlation length of the disorder potential.

\noindent Note that due to translational symmetry of the disorder correlation function \eqref{eq:disorder_corr}, the disorder-averaged spectral function $\langle S_{0X}(\varepsilon)\rangle_s$ will not depend on $X$, so we drop this dependence in subsequent equations. Calculating the disorder averaged of the square of transition matrix element in Eq. \eqref{eq:semi_master}, we obtain 

\begin{equation}
\sigma_{xx}=4 e^2\frac{1}{V}\sum_{X}\frac{1}{V}\sum_{\bm{q}}\left(-\frac{\partial f}{\partial \varepsilon}\right)\frac{(q_yl^2)^2}{2}2\pi \mathcal{F}_{\bm{q}}e^{-q^2l^2/2}\langle S_{0}(\varepsilon)\rangle_s \langle S_{0}(\varepsilon)\rangle_s.
\end{equation}

\noindent Taking into account the rotational symmetry of $\mathcal{F}_{\bm{q}}$ let us replace $q_y^2$ with $q^2/2$. Also replacing summations with respect to quantum numbers with integrals according to the rules

\be
\sum_{\bm{q}}\dots\to V \int\frac{d^2q}{(2\pi)^2}\dots; \qquad \sum_{X}\dots\to\frac{L_y}{2\pi l^2}\int_0^{L_x}dX\dots
\ee

\noindent and performing the integration over $X$ we obtain

\begin{equation}\label{eq:almost_there}
\sigma_{xx}=e^2l^2\int d\varepsilon\left(-\frac{\partial f}{\partial \varepsilon}\right)\langle S_0(\varepsilon)\rangle_s^2\int\frac{d^2q}{(2\pi)^2}\mathcal{F}_{\bm{q}}q^2e^{-q^2l^2/2}.
\end{equation}
\begin{figure}[t]
\includegraphics[scale=1]{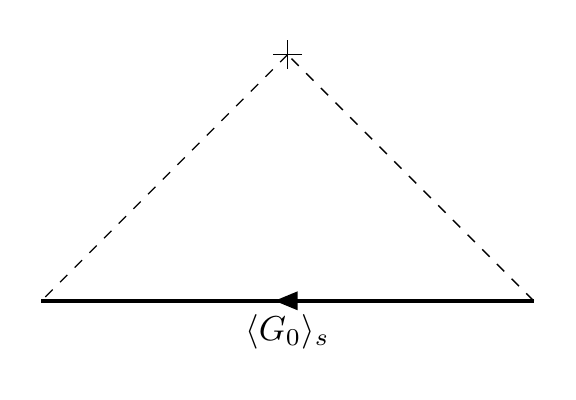}
\caption{\label{fig:scba} Self-energy of the zeroth Landau level in the self-consistent Born approximation}
\end{figure}
\noindent Because the unperturbed Landau levels have zero width, the broadening must be calculated self-consistently. The self-energy $\langle \Sigma_0(\varepsilon)\rangle_s$ is thus defined via the self-consistent equations (see Fig. \ref{fig:scba})

\begin{eqnarray}\label{eq:dis_green}
\langle\Sigma_0(\varepsilon)\rangle_s &=& \frac{\gamma^2}{4} \langle G_{0}(\varepsilon)\rangle_s,
\nonumber\\
\langle G_{0}(\varepsilon)\rangle_s&=&\frac{1}{\varepsilon-\langle\Sigma_0(\varepsilon)\rangle_s},
\end{eqnarray}

\noindent where $\langle G_0(\varepsilon)\rangle_s$ is the disorder-averaged zeroth Landau level Green function and
\begin{equation}
\frac{\gamma^2}{4}=\int\frac{d^2q}{(2\pi)^2}\mathcal{F}_{\bm{q}}e^{-q^2l^2/2}.
\end{equation}
In Eq.~(\ref{eq:dis_green}) we have neglected diagrams with self-intersecting impurity lines (the resulting approximation is called the self-consistent Born approximation). Thus, the following equation follows 

\begin{equation}\label{eq:quadratic}
\langle\Sigma_0(\varepsilon)\rangle_s=\frac{\gamma^2}{4}\frac{1}{\varepsilon-\langle\Sigma_0(\varepsilon)\rangle_s},
\end{equation}

\noindent The quadratic equation \eqref{eq:quadratic} has the following solution

\begin{equation}\label{eq:solution}
\langle \Sigma_0(\varepsilon)\rangle_s=\frac{\varepsilon}{2}-i\sqrt{\frac{\gamma^2}{4}-\frac{\varepsilon^2}{4}}.
\end{equation}

\noindent From this, one can see that $\gamma$ is the width of the zeroth Landau level.  Taking the imaginary part of the disorder averaged Green function given by Eqs. \eqref{eq:dis_green} and \eqref{eq:solution}, we obtain 

\begin{equation}
\langle S_0(\varepsilon)\rangle_s^2=\frac{4}{\pi^2\gamma^2}\left(1-\frac{\varepsilon^2}{\gamma^2}\right).
\end{equation}

\noindent Plugging this into Eq. \eqref{eq:almost_there}, we obtain

\begin{equation}
\sigma_{xx}=\frac{4e^2l^2}{\pi^2\gamma^2}\int d\varepsilon\left(-\frac{\partial f}{\partial \varepsilon}\right)\left(1-\frac{\varepsilon^2}{\gamma^2}\right)\int\frac{d^2q}{(2\pi)^2}\mathcal{F}_{\bm{q}}q^2e^{-q^2l^2/2}.
\end{equation}

\noindent Introducing the transport scattering rate $\gamma_{tr}$ as 

\begin{equation}
\frac{\gamma_{tr}^2}{4}=\int \frac{d^2q}{(2\pi)^2}\mathcal{F}_{\bm{q}}l^2q^2e^{-q^2l^2/2},
\end{equation}
\noindent one can rewrite this equation as

\begin{equation} \label{eq:sigma_xx_semifinal}
\sigma_{xx}=\frac{e^2}{\pi^2}\int d\varepsilon \left(-\frac{\partial f}{\partial \varepsilon}\right)\left(1-\frac{\varepsilon^2}{\gamma^2}\right)\frac{\gamma_{tr}^2}{\gamma^2}.
\end{equation}
 
\noindent Using the disorder correlation function \eqref{eq:disorder_corr},  and taking the limit of extremely large correlation length $\xi\gg l$, we obtain

\begin{equation}
\frac{\gamma_{tr}^2}{\gamma^2}=\frac{2 l^2}{\xi^2}.
\end{equation}

\noindent The integral  in Eq.~(\ref{eq:sigma_xx_semifinal}) with respect to $\varepsilon$ depends only weakly on temperature if $\gamma>kT$. Under this assumption we obtain (restoring $\hbar$)

\begin{equation}
\sigma_{xx}=\frac{e^2}{\pi^2\hbar}\frac{2l^2}{\xi^2}\propto \frac{1}{H}.
\end{equation}

\section{Equivalence of Feynman diagrams and center migration equations}

In this section we demonstrate that the center migration theory is equivalent to the usual Feynman diagrams. We will do this for the three-dimensional model and the generalization to the two dimensional case will be obvious. The treatment here will be very close to that of Ref. \cite{Ando}. 

In linear response theory the static conductivity tensor is given by the equation

\be 
\sigma_{ik}=-\lim_{\omega\to 0}\frac{\Im Q^{R}_{ik}(\omega)}{\omega},
\ee

\noindent where $Q^R_{ik}(\omega)$ is the Fourier transform of the current-current linear response function

\be
Q^R_{ik}(\omega)=-iV^{-1}\int_{0}^{\infty}dt e^{i\omega t}\langle [j_i(t), j_k(0)]\rangle,
\ee

\noindent where $j_{i}$ is the many-body current operator and the brackets $\langle\dots\rangle$ stand for thermal averaging. The function $Q_{ik}^R(\omega)$ can in turn be obtained by analytic continuation from discrete imaginary frequencies of  $Q_{ik}(i\omega_m)$, which is the Fourier transform of the imaginary time current-current correlation function,

\be\label{eq:curr_curr}
Q_{ik}(i\omega_m)=-V^{-1}\int_{0}^{1/T}d\tau e^{i\omega_m\tau}\langle  T_{\tau}j_{i}(\tau)j_{k}(0)\rangle,
\ee

\noindent where $\omega_m=2\pi m T$ with $m$ integer, and $T_{\tau}$ stands for time ordering in imaginary time.  The function $Q_{ik}(i\omega_m)$ can be represented as a sum of connected diagrams with the two operator insertions $j_i=ev\int d\bm{r}\psi^\dag \sigma_i\psi$ and $j_k=ev\int d\bm{r}\psi^\dag \sigma_k \psi$, where $\psi$ is the electron field operator.

\begin{figure}[t]
\includegraphics[scale=0.71]{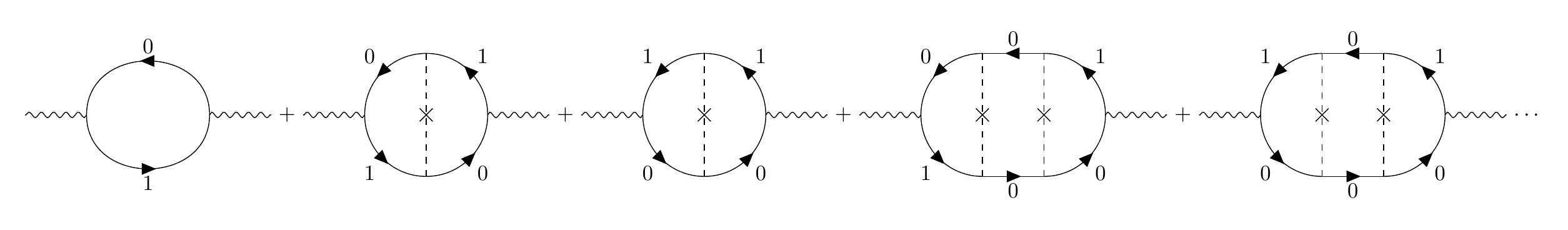} 
\caption{\label{fig:cond_diagr}
The ladder series of Feynman diagrams contributing to conductivity to first nontrivial order in powers of $\omega_c^{-1}$. The labels on the internal lines display the absolute value of the Landau level number}
\end{figure}

The diagrams that contribute to conductivity $\sigma_{xx}$ in the first Born approximation and to the first non-trivial order in powers of $\omega_c^{-1}$ are depicted in Fig. \ref{fig:cond_diagr}. The Green functions corresponding to fermion lines are disorder averaged and because, as we will see, disorder couples Landau levels with equal but opposite Landau level numbers, each line is labelled not by the Landau level number but by the absolute Landau level number.  The reasoning behind this choice of diagrams is as follows.  Because the chemical potential is zero and the temperature is low such that $kT \ll \omega_c$, the diagrams that give the largest contribution must have as many Landau level zero internal lines as possible. Each internal line that is other than  Landau level zero brings in an additional power of $\omega_c^{-1}$. On the other hand, since the current operator $\sigma_{x}$ couples Landau levels with numbers differing by $\pm 1$, some of the states propagating along internal lines must have absolute Landau level number equal to one.

\begin{figure}[t]
\begin{tabular}{cc}
\begin{overpic}[width=0.4\linewidth]{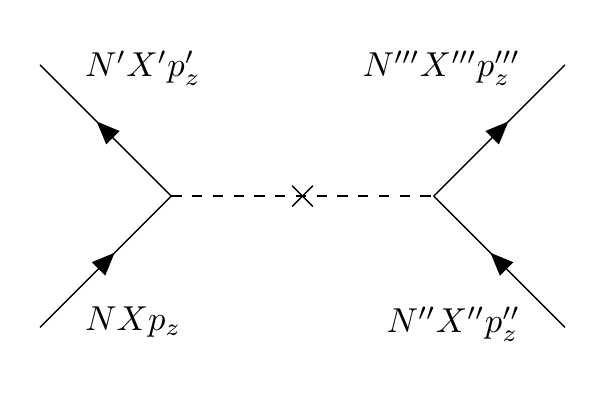}\put(-5,85){(a)}\end{overpic}&
\begin{overpic}[width=0.3\linewidth]{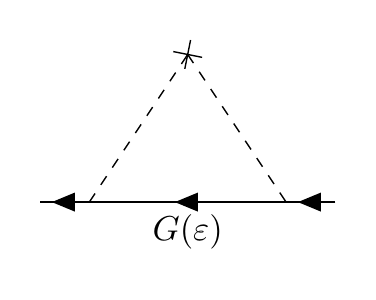}\put(-5,85){(b)}\end{overpic}
\end{tabular}
\caption{\label{fig:proper} Panel (a) The proper four-fermion vertex part to first Born approximation. Panel (b) Self-energy operator in the first (self-consistent) Born approximation}
\end{figure}

The proper four-fermion vertex part used in the first Born approximation, depicted in Fig. \ref{fig:proper} (a), equals

\be \label{eq:vertex}
\Big\langle \langle N'X'p_z'|U|NXp_z  \rangle \langle N'''X'''p_z'''|U|N''X''p_z''\rangle\Big\rangle_s=\frac{1}{V}\sum_{\bm{q}}N_s|u_{\bm{q}}|^2\langle N'X'p_z'|e^{i\bm{q}\bm{r}}|NXp_z  \rangle \langle N'''X'''p_z'''|e^{-i\bm{q}\bm{r}}|N''X''p_z''\rangle,\nonumber\\
\ee

\noindent where

\be\label{eq:nonzero}
\langle N'X'p_z'|e^{i\bm{q}\bm{r}}|NXp_z  \rangle={}&&\delta_{X', X+q_yl^2}\delta_{p_z', p_z+q_z}\exp{\bigg[iq_x\frac{(X+X')}{2}+i\varphi(|N'|-|N|)\bigg]}\nonumber\\&&{}\times \bigg\{\frac{1}{2}\left(1+\frac{vp_z'}{E_{N'}}\right)^{1/2}\left(1+\frac{vp_z}{E_{N}}\right)^{1/2}L_{|N'|,|N|}\left(\frac{q_\perp l}{\sqrt{2}}\right)\nonumber\\
&&\qquad{}+\frac{1}{2}\mbox{sign}(N')\mbox{sign}(N)\left(1-\frac{vp_z'}{E_{N'}}\right)^{1/2}\left(1-\frac{vp_z}{E_{N}}\right)^{1/2}L_{|N'|-1,|N|-1}\left(\frac{q_\perp l}{\sqrt{2}}\right)\bigg\}\nonumber\\
\ee

\noindent for $N\neq 0$ and $N'\neq 0$, with the polar angle $\varphi=\mbox{arg}(q_x+iq_y)$ and 

\be
L_{n', n}(x)=L_{n,n'}(x)=\sqrt{\frac{n!}{n'!}}{(ix)}^{n'-n}L_{n}^{n'-n}(x^2) e^{-x^2},
\ee

\noindent where $L_{n}^{n'-n}$ is the generalized Laguerre polynomial of degree $n$.  Similarly,

\be\label{eq:zero}
\langle 0 X'p_z'|e^{i\bm{q}\bm{r}}|NXp_z  \rangle=\frac{1}{\sqrt{2}} \delta_{X', X+q_yl^2}\delta_{p_z', p_z+q_z}\exp{\bigg[iq_x\frac{(X+X')}{2}-i\varphi |N|\bigg]}\left(1+\frac{vp_z}{E_{N}}\right)^{1/2}L_{0,|N|}.
\ee

\noindent Substitution of Eq. \eqref{eq:nonzero} or Eq. \eqref{eq:zero} into Eq. \eqref{eq:vertex} and subsequent integration over the polar angle $\varphi$ (assuming spherical symmetry of $u_{\bm{q}}$) makes the proper vertex proportional to $\delta_{|N|+|N''|,|N'|+|N'''|}$  at $X=X'''$, that is, at $X=X'''$ the four-fermion vertex conserves the absolute Landau level number. For that reason, diagrams in the ladder series depicted in Fig. \ref{fig:cond_diagr} starting from the fourth one all vanish. For the same reason, also the second diagram vanishes. So, one only needs to evaluate the first and the third  diagrams in Fig.~\ref{fig:cond_diagr}.  In what follows, we show that also the third diagram vanishes to the first nontrivial order in powers of $\omega_c^{-1}$. The first one has already been evaluated in Ref. \citep{Abrikosov}. For the sake of completeness, we reiterate this calculation here too. 

Everywhere below we will suppress the disorder averaging brackets in the notation for the Green functions assuming they are already disorder averaged, {\it i.e.},

\be
G(\varepsilon)=\bigg\langle\frac{1}{\varepsilon - {\cal H}}\bigg\rangle_s,
\ee
 
\noindent where ${\cal H}=v\bm{\sigma}(\bm{p}-(e/c)\bm{A})+U={\cal H}_0+U$, with $U$ the disorder potential. Defining the self-energy operator $\Sigma(\varepsilon)$ via the equation $G(\varepsilon)=(\varepsilon-{\cal H}_0-\Sigma(\varepsilon))^{-1}$, 
 within the self-consistent Born approximation the self-energy is given by the diagram in Fig. \ref{fig:proper} (b), {\it i.e.}

\be\label{eq:SE}
\Sigma(\varepsilon)=\langle UG(\varepsilon)U\rangle_s.
\ee

\noindent Due to the uniformity of distribution of scattering centers,  the matrix elements of $\Sigma(\varepsilon)$ [and, consequently, of $G(\varepsilon)$] will be diagonal in quantum numbers $X$, $p_z$ and will be independent of $X$. Within the self-consistent Born approximation, due to the conservation of the absolute Landau level number by the proper four-fermion vertex, $\Sigma(\varepsilon)$ will also be diagonal in $|N|$, but not necessarily in $N$ itself.   For Landau levels $\pm 1$ let us introduce a $2\times 2$ matrix $G_{1p_z}(\varepsilon)$ whose matrix elements are defined as

\be 
[G_{1p_z}(\varepsilon)]_{\sigma\sigma'}=\bigg\langle\Big\langle\sigma X p_z\Big|\frac{1}{\varepsilon - {\cal H}}\Big|\sigma' X p_z\Big\rangle\bigg\rangle_s.
\ee

\noindent Here, $\sigma$ and $\sigma'$ are equal to $\pm 1$, the internal brackets stand for quantum mechanical averaging and the external for disorder averaging. The corresponding self-energy matrix $\Sigma_{1p_z}(\varepsilon)$ is defined such that $G_{1p_z}=[\varepsilon-E_{1}(p_z)\sigma_z-\Sigma_{1p_z}]^{-1}$ [here, $E_{1}(p_z)\sigma_z$ is a matrix with the unperturbed energy values of Landau levels $\pm 1$ on the diagonal]. Also for the zeroth Landau level introduce

\be
G_{0p_z}=\bigg\langle\Big\langle 0 X p_z\Big|\frac{1}{\varepsilon - {\cal H}}\Big|0 X p_z\Big\rangle\bigg\rangle_s.
\ee

\noindent To lowest nontrivial order in powers of $\omega_c^{-1}$ the equation for $\Sigma_{1p_z}$, which is just the projection of Eq. \eqref{eq:SE} on Landau levels $\pm 1$,  has the form,
 
\be \label{eq:SE_firstLL}  
[\Sigma_{1p_z}(\varepsilon)]_{\sigma\sigma'}=\sum_{X'p_z'}\Big\langle\langle\sigma Xp_z|U|0X'p_z'\rangle \langle\ 0 X'p'_z|U|\sigma' Xp_z\rangle\Big\rangle_s G_{0p_z'}(\varepsilon).
\ee

Now we have all the ingredients necessary for the calculation of $\sigma_{xx}$. The contribution of the first diagram in Fig.~\ref{fig:cond_diagr} to $\sigma_{xx}$ equals

\be
\Delta \sigma_{xx}^{(1)}=4 v^2e^2 V^{-1}\int\frac{d\varepsilon}{2\pi}\left(-\frac{\partial f}{\partial \varepsilon}\right)\mbox{tr}\Big[\sigma_x \Im G(\varepsilon+i0)\sigma_x \Im G(\varepsilon+i0)\Big]_{O(\omega_c^{-2})},
\ee

\noindent where $\Im G(\varepsilon+i0)=(1/2i)[G(\varepsilon+i0)-G(\varepsilon-i0)]$ and  the subscript $O(\omega_c^{-2})$ means that only the contribution of order $\omega_c^{-2}$ needs to be extracted.
A factor of 2 coming from spin degeneracy is also included. The evaluation of this term can be done by inserting the resolution of identity twice into this equation and keeping only the projectors on Landau levels $0$ and $\pm 1$. Keeping in mind that $\langle \sigma X p_z|\sigma_x|0 X p_z\rangle = (i/\sqrt{2}) \sigma (1-vp_z/E_{\sigma})^{1/2}$ and neglecting $vp_z/E_{\sigma}$ we obtain

\be
\Delta \sigma_{xx}^{(1)}=4v^2e^2V^{-1}\int\frac{d\varepsilon}{2\pi}\left(-\frac{\partial f}{\partial \varepsilon}\right)\sum_{\sigma\sigma'}\sum_{Xp_z}\sum_{X'p_z'}\sigma\sigma' &&\Im G_{0p_z}(\varepsilon+i0)\big[\Im G_{1p_z'}(\varepsilon+i0)\big]_{\sigma\sigma'}.
\ee

\noindent Taking into account that

\be\label{eq:Greenonepz}
 G_{1p_z'}(\varepsilon)=(\varepsilon-\sigma_z E_{1}(p_z'))^{-1}+(\varepsilon-\sigma_z E_{1}(p_z'))^{-1}\Sigma_{1p_z}(\varepsilon)(\varepsilon-\sigma_z E_1(p_z'))^{-1}+\dots,
\ee

\noindent the imaginary part $\Im G_{1p_z'}(\varepsilon+i0)=\omega_c^{-2}\sigma_z\Im \Sigma_{1p_z'}(\varepsilon+i0)\sigma_z$ to order $\omega_c^{-2}$.  Using Eq. \eqref{eq:SE_firstLL} we obtain

\be\label{eq:no_vertex_corr}
\Delta \sigma_{xx}^{(1)}=(4v^2e^2/\omega_c^2)V^{-1}\int\frac{d\varepsilon}{2\pi}\left(-\frac{\partial f}{\partial \varepsilon}\right)\sum_{\sigma\sigma'}\sum_{Xp_z}\sum_{X'p_z'}&&\Im G_{0p_z}(\varepsilon+i0)\Im G_{0p_z'}(\varepsilon+i0)\nonumber\\&&{}\times\Big\langle\langle\sigma Xp_z|U|0X'p_z'\rangle \langle\ 0 X'p'_z|U|\sigma' Xp_z\rangle\Big\rangle_s.\nonumber\\
\ee

The contribution to $\sigma_{xx}$ of the second and third diagrams in Fig.  \ref{fig:cond_diagr} is given by the expression (with spin degeneracy taken into account)

\be
\Delta\sigma_{xx}^{(2,3)}=2 e^2 v^2V^{-1}\int\frac{d\varepsilon}{(2\pi)}\left(-\frac{\partial f}{\partial \varepsilon}\right)\bigg[ \mbox{Re}\,\mbox{tr}\Big\langle &&\sigma_x G(\varepsilon+i0)UG(\varepsilon+i0)\sigma_x\nonumber\\&&{}\times\Big[G(\varepsilon-i0)UG(\varepsilon-i0)-G(\varepsilon+i0)UG(\varepsilon+i0)\Big]\Big\rangle_{s}\bigg]_{O(\omega_c^{-2})}.\nonumber\\
\ee

\noindent  We now insert resolutions of identity and  keep only projectors on Landau levels $0$ and $\pm 1$.  Furthermore, we neglect $vp_z/E_{\pm 1}$ in the matrix elements of $\sigma_x$, while keeping only the first term in Eq. \eqref{eq:Greenonepz}.  The latter is enough because there must always be two $G_{1p_z}$ lines, see Fig. \eqref{fig:cond_diagr}. We obtain 

\be\label{eq:vertex_corr}
\Delta \sigma_{xx}^{(2,3)}=-(2e^2v^2/\omega_c^2)V^{-1}\int\frac{d\varepsilon}{2\pi}\left(-\frac{\partial f}{\partial \varepsilon}\right)\sum_{\sigma\sigma'}\sum_{XX'}\sum_{p_zp_z'}&&\Im G_{0p_z}(\varepsilon+i0)\Im G_{0p_z'}(\varepsilon+i0)\nonumber\\&&{}\times\Big\langle\langle\sigma'X'p_z'|U|0Xp_z\rangle\langle \sigma X p_z|U|0X'p_z'\rangle+ \mbox{c.c.}\Big\rangle_s
\ee

\noindent where $\mbox{c.c.}$ stands for complex conjugate. Note that after disorder averaging this contribution vanishes because of the conservation of the absolute Landau level number by the proper four-fermion vertex. Such vanishing of the vertex corrections is a special property of the zeroth Landau level.

To show the equivalence between the Feynman diagram calculation and the center migration theory, we take the sum of Eqs. \eqref{eq:no_vertex_corr} and \eqref{eq:vertex_corr} before disorder averaging. Then, we introduce the ladder operator $a=(l^{-1}(x-X)+ilp_x)/{\sqrt{2}}$, which lowers the absolute Landau level number and annihilates the zeroth Landau level. Note that, neglecting $vp_z/E_{\pm 1}$, we can write for any $\sigma=\pm 1$ 

\be 
\langle \sigma X p_z|U|0X'p_z'\rangle=\langle 0 Xp_z| \frac{a}{\sqrt{2}}U|0X'p_z'\rangle
\ee

\noindent and similarly for complex conjugated matrix elements. Then the sum of Eqs. \eqref{eq:no_vertex_corr} and \eqref{eq:vertex_corr} can be written as

\be
\sigma_{xx}=(8e^2v^2/\omega_c^2)V^{-1}\int\frac{d\varepsilon}{2\pi}\left(-\frac{\partial f}{\partial \varepsilon}\right)\sum_{Xp_zX'p_z'}\Im G_{0p_z}(\varepsilon+i0)\Im G_{0p_z'}(\varepsilon+i0)\bigg\langle\Big|\Big\langle 0Xp_z\Big|\left[\frac{a+a^\dag}{\sqrt{2}}, U\right]\Big|0X'p_z'\Big\rangle\Big|^2\bigg\rangle_s.\nonumber\\
\ee

\noindent Noting that $(a+a^\dag)/{\sqrt{2}}=(x-X)/l$, where $X=l^2p_y$ is the center position operator, also that $\Im G_{0p_z}(\varepsilon+i0)=-\pi\delta(\varepsilon-E_{0}(p_z))$ and $\omega_c=\sqrt{2}v/l$, and integrating with respect to $\varepsilon$, we obtain Eq. \eqref{eq:master}.

\section*{Conclusion}

In this paper we presented a simpler derivation of Abrikosov's quantum magnetoresistance, based on the picture of diffusing cyclotron centers. It is better suited for the calculation of magnetoresistance in the extreme quantum limit because it admits a natural restriction of dynamics to a single Landau level, is very intuitive and allows the whole calculation to be reduced just to single application of the Fermi golden rule. Furthermore, in a random potential smooth on the scale of the magnetic length, the cyclotron centers drift according to classical laws, which together with the linear-in-$H$ density of states leads to a linear magnetoresistance for a compensated Weyl semimetal in three dimensions. 

Another advantage of this approach is that it works equally well for a two-dimensional model too. In particular, for undoped graphene, a smooth random potential will also give rise to a linear magnetoresistance.

\section*{Acknowledgments} A.P. acknowledges support from the European Commission under the EU Horizon 2020 MSCA-RISE-2019 programme (project 873028 HYDROTRONICS). A.P. and A.K. acknowledge support from the Leverhulme Trust under the grant RPG-2019-363.   
\bibliographystyle{apsrev4-2}
\bibliography{ref}

\end{document}